%% file: main.tex
\documentclass[sigconf]{acmart}
\makeatletter
\def\@copyrightpermission{%
  \ifcase\acm@copyrightmode\relax
  \or
  \or
  \or
  \or
  \or
  \or
  \or
  \or
  \or
  \or
  \or
  \or
  \or
  \or
  \or
  \or
  \fi
}

\makeatother

\acmConference[Q-SET '25]{International Workshop on Quantum Software Engineering and Technology}{Aug 31 –- Sep 5, 2025}{USA}
\setcopyright{none}
\renewcommand\footnotetextcopyrightpermission[1]{}

\usepackage{lipsum}  
\usepackage{color,soul}
\usepackage{tabularx}
\usepackage{mfirstuc}
\usepackage{tabularx}
\usepackage{ragged2e} 
\usepackage{lipsum} 
\usepackage{braket}
\usepackage{amsmath}
\usepackage{tikz}
\usetikzlibrary{3d,arrows.meta}

\usepackage[acronym]{glossaries}
\makeglossaries
\input{abbreviations}

\begin{document}

\title{Toward a Brazilian Research Agenda in Quantum Software Engineering: A Systematic Mapping Study}




\author{Filipe Fernandes}
\affiliation{%
 \institution{Federal Institute of Southeast of Minas Gerais\\(IF Sudeste MG)}
 \city{Manhuaçu}
 \state{Minas Gerais}
 \country{Brazil}}
 \email{filipe.fernandes@ifsudestemg.edu.br}

\author{Cláudia Werner}
\affiliation{%
 \institution{Federal University of Rio de Janeiro\\(COPPE/UFRJ)}
 \city{Rio de Janeiro}
 \state{Rio de Janeiro}
 \country{Brazil}}
 \email{werner@cos.ufrj.br}




\renewcommand{\shortauthors}{Fernandes and Werner}

\input{newcommands}

\begin{abstract} 
\textit{Context}: Quantum Software Engineering (QSE) has emerged as a promising discipline to support the development of quantum applications by integrating quantum computing principles with established software engineering practices.
\textit{Problem}: Despite recent growth, QSE still lacks standardized methodologies, tools, and guidelines, and countries like Brazil have minimal representation in its development.
\textit{Objective}: This study aims to map the current state of QSE, identifying research trends, contributions, and gaps to support future investigations and initiatives.
\textit{Methodology}: A systematic mapping study was conducted across major databases, retrieving \totalretrieved~ studies, of which \totalexcluded~ were excluded, resulting in \totalfinalstudies~ studies for analysis. Publications were classified by study type, research type, and SWEBOK alignment.
\textit{Results}: The majority studies focused on \textit{Software Engineering Models and Methods}, \textit{Software Architecture}, and \textit{Software Testing}. Conceptual and technical proposals dominate, while empirical validations are still few.
\textit{Conclusions}: QSE remains a maturing field. Standardization, empirical research, and the inclusion of developing countries are essential. As a key contribution, this study proposes a Brazilian Research Agenda in QSE to guide national efforts and foster a strong scientific community.
\end{abstract}





\settopmatter{printacmref=false} 
\maketitle

\input{sections/Introduction}
\input{sections/Background}
\input{sections/RelatedWork}
\input{sections/ResearchMethod}
\input{sections/Results}

\input{sections/Discussion}
\input{sections/Agenda}

\input{sections/Conclusion}

\section*{Artifact Availability}
Data used in the analysis, as well as supplementary materials can be accessed in \href{\repository}{\repository}.

\bibliographystyle{ACM-Reference-Format}
\bibliography{references}

\end{document}

%% file: abbreviations.tex
\newacronym{SE}{SE}{Software Engineering}
\newacronym{QSE}{QSE}{Quantum Software Engineering}
\newacronym{SWEBOK}{SWEBOK}{Software Engineering Body of Knowledge}
\newacronym{IBM}{IBM}{International Business Machines}
\newacronym{RQS}{RQs}{Research Questions}
\newacronym{DSL}{DSL}{Domain-Specific Language}
\newacronym{SMS}{SMS}{Systematic Mapping Study}
\newacronym{QC}{QC}{Quantum Computing}
\newacronym{UN}{UN}{United Nations}

%% file: newcommands.tex
\newcommand{\rqzero}{What is the state of the art and evolution of \gls{QSE}}

\newcommand{\rqone}{What are the bibliometric and contextual characteristics of \gls{QSE} publications?}
\newcommand{\rqonecode}{RQ1}

\newcommand{\rqtwo}{What are the main contributions, challenges, and research gaps addressed by \gls{QSE} studies?}
\newcommand{\rqtwocode}{RQ2}


\newcommand{\rqoneone}{What is the temporal distribution of \gls{QSE} publications?} 
\newcommand{\rqoneonecode}{RQ1}

\newcommand{\rqonetwo}{Who are the most active institutions and authors in the field?}  
\newcommand{\rqonetwocode}{RQ2}

\newcommand{\rqonethree}{Which countries are leading \gls{QSE} research?} 
\newcommand{\rqonethreecode}{RQ3}

\newcommand{\rqonefour}{What are the most frequent publication venues?} 
\newcommand{\rqonefourcode}{RQ4}

\newcommand{\rqonefive}{Which research strategies are most prevalent?} 
\newcommand{\rqonefivecode}{RQ5}



\newcommand{\rqtwoone}{Which areas of SWEBOK are most frequently addressed in \gls{QSE} research?} 
\newcommand{\rqtwoonecode}{RQ6}





\newcommand{\blochsphere}[3]{
  \begin{scope}[shift={#3}, scale=0.7]
    \draw[thick] (0,0) circle (1); 
    \draw[dashed] (0,0) ellipse (1 and 0.3); 
    \draw[thick,->,blue] (0,0) -- #2; 
    \node[above] at (0,1.1) {\(\ket{0}\)};
    \node[below] at (0,-1.1) {\(\ket{1}\)};
    \node[below=8pt] at (0,-1.6) {\(#1\)};
  \end{scope}
}

\newcommand{\repository}{https://bit.ly/BRAgendaQSE}

\newcommand{\acmretrieved}{11}
\newcommand{\acmexcluded}{10}
\newcommand{\acmaccepted}{1}

\newcommand{\ieeeretrieved}{112}
\newcommand{\ieeeexcluded}{112}
\newcommand{\ieeeaccepted}{0}

\newcommand{\wosretrieved}{1715}
\newcommand{\wosexcluded}{1626}
\newcommand{\wosaccepted}{89}

\newcommand{\scpretrieved}{1381}
\newcommand{\scpexcluded}{1304}
\newcommand{\scpaccepted}{77}

\newcommand{\totalretrieved}{3219}
\newcommand{\totalexcluded}{3052}
\newcommand{\totalfinalstudies}{167}

%% file: sections/Introduction.tex
\section{Introduction} \label{ref:introduction}

\gls{UN} has declared 2025 the International Year of Quantum Science and Technology, encouraging nations and institutions to engage in scientific, educational, and technological initiatives that promote quantum innovation \cite{quantum2025}. This global mobilization reflects the increasing relevance of \gls{QC}, a computational paradigm that leverages quantum-mechanical phenomena, such as superposition, entanglement, and interference, to solve problems that are intractable for classical computers.

\gls{QC} promises significant advancements in fields such as cryptography, optimization, materials science, and drug discovery. However, harnessing its potential requires overcoming complex challenges, particularly in the domain of software development. Quantum software does not follow the same logic, abstractions, or tooling as classical software, demanding novel models, languages, and engineering practices \cite{hughes2021quantum}.

In this context, the emerging field of \gls{QSE} has gained attention as a research frontier that aims to systematize the development, testing, debugging, and maintenance of quantum software \cite{ali2022software}. Additionally, \gls{QSE} landscape remains fragmented and lacks a consolidated research agenda, especially in regions like Latin America \cite{Aparicio-Morales2024}. Therefore, there is a pressing need to understand the current state of \gls{QSE} research.

In order to address these questions, this work conducts a \gls{SMS} of the literature in \gls{QSE}. The goal is not only to contribute to a comprehensive understanding of the field, but also to lay the foundation for the development of a Brazilian research agenda in \gls{QSE}. Such an agenda is critical to guide national efforts, support funding strategies, and stimulate academic and industrial collaboration in quantum technologies.


%% file: sections/Background.tex
\section{Theoretical Background} \label{sec:background}


\subsection{Quantum Computing} \label{sec:quantumcomputing}
In classical computers, information is represented as the binary digits $0$ or $1$. These are called bits. 
Quantum bits or qubits are similar to bits in that there are two measurable states called the $0$ and $1$ states. 
However, unlike classical bits, qubits can also be in a superposition state of these $0$ and $1$ states \cite{hughes2021quantum}.

Superposition is one of the fundamental aspects of quantum computing, because it enables an operation to simultaneously affect all amplitudes, a phenomenon known as quantum parallelism \cite{hughes2021quantum}.
The state of a qubit is enclosed in the right half of an angled bracket, called the ``\textit{ket}". 
A qubit, $\ket{\psi}$, could be in a $\ket{0}$ \textbf{or} $\ket{1}$ state or even a superposition of both $\ket{0}$ \textbf{and} $\ket{1}$. This is written as 

\begin{equation}
    \ket{\psi} = \alpha\ket{0} + \beta\ket{1} \text{,}
\end{equation}

where $\alpha$ and $\beta$ are complex numbers known as amplitudes that satisfy

\begin{equation}
    |a|^2 + |b|^2 = 1 \text{.}
\end{equation}

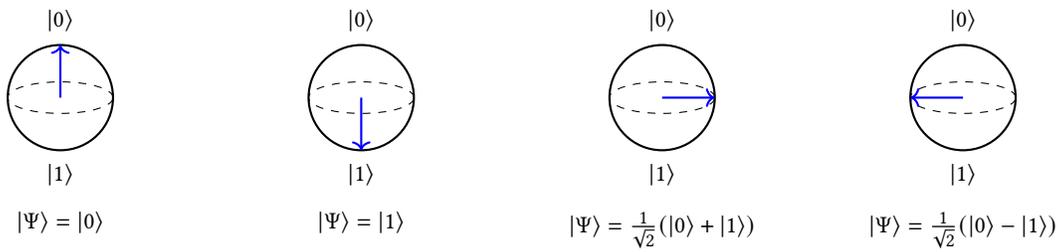
\begin{figure*}[ht]
\centering
\begin{tikzpicture}
  \blochsphere{\ket{\Psi} = \ket{0}}{(0,1)}{(-6,0)}
  \blochsphere{\ket{\Psi} = \ket{1}}{(0,-1)}{(-2,0)}
  \blochsphere{\ket{\Psi} = \frac{1}{\sqrt{2}}(\ket{0} + \ket{1})}{(1,0)}{(2,0)}
  \blochsphere{\ket{\Psi} = \frac{1}{\sqrt{2}}(\ket{0} - \ket{1})}{(-1,0)}{(6,0)}
\end{tikzpicture}
\caption{Representation of different quantum states of a qubit in the Bloch sphere \cite{hughes2021quantum}}
\label{fig:blochsphere}
\end{figure*}

The representation of a qubit can be modeled geometrically using the Bloch sphere, a visual tool analogous to the unit circle in trigonometry. Each point on the surface of the sphere represents a possible superposition state of the qubit, with the north and south poles associated with the computational states $\ket{0}$ and $\ket{1}$, respectively \cite{hughes2021quantum}.
The state vector of the qubit is represented by an arrow pointing to a specific position on the sphere. Superposition states correspond to intermediate positions, especially at the equator, where the measurement results in either $\ket{0}$ or $\ket{1}$ with equal probabilities. Changes in the state of the qubit are reflected by rotations of this vector about the sphere, allowing the dynamic visualization of quantum evolution. Figure~\ref{fig:blochsphere} illustrates typical examples of quantum states visualized on the Bloch sphere.

  

Another fundamental concept is quantum entanglement. When two or more qubits are entangled, the state of one directly influences the other, even if they are physically separated \cite{hughes2021quantum}. This property is essential for quantum communication and cryptography algorithms, such as
Quantum Teleportation \cite{QuantumTeleportation}, Superdense Coding \cite{SuperdenseCoding}, Quantum One-Time Pad \cite{QuantumOneTimePad}, among others.

\subsection{Development of Quantum Software} \label{sec:quantumsoftware}
Since \gls{QC} is a new paradigm in information processing based on the fundamentals of quantum mechanics (such as superposition and entanglement), the development of software for \gls{QC} requires new programming languages, execution models and specific tools that enable the creation, simulation and execution of quantum algorithms \cite{ali2022software}.

This development can occur in two main models: quantum gate-based computing, widely used in frameworks such as Qiskit\footnote{\href{https://www.ibm.com/quantum/qiskit}{https://www.ibm.com/quantum/qiskit}}, and adiabatic computing, common in optimization problems and implemented, for example, in architectures such as D-Wave\footnote{\href{https://www.dwavequantum.com/}{https://www.dwavequantum.com/}}.

A quantum circuit is a set of quantum gates that are applied to a group of qubits.
Analogous to the classical computing model, information processing is performed through logic gates that operate on qubits, known as quantum gates.
Pauli's gates ($X$, $Y$, and $Z$), Hadamard ($H$), and Controlled-NOT ($CNOT$), are some examples of quantum gates used in quantum circuits to transform the states of qubits over time.

\begin{figure}
    \centering
    \includegraphics[width=1\linewidth]{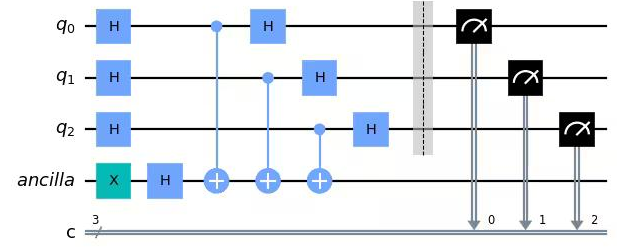}
    \caption{Quantum circuit implementation \cite{Qiskit}}
    \label{fig:quantumcircuit}
\end{figure}

These transformations in a quantum circuit are based on linear algebra, using unitary vectors and matrices to represent and manipulate quantum states. This differs from classical computing, which is based on Boolean algebra, where operations such as $AND$, $OR$, and $NOT$ operate on bits. In quantum circuits, quantum logic gates operate on qubits. Figure~\ref{fig:quantumcircuit} shows an example of a quantum circuit.

The second approach, quantum annealing, is used in optimization problems. In this model, the computer slowly evolves the initial quantum state to find the lowest energy state of a system. This technique is the basis for the operation of D-Wave computers and has been applied in areas such as logistics, machine learning and combinatorial problems \cite{rajak2023quantum}.

Although they are fundamentally different, both approaches rely on specific tools for the development, simulation and validation of algorithms.
Furthermore, as it is an area still undergoing consolidation, the development of quantum software involves not only coding, but also the need for specialized modeling, testing and debugging, as well as experimental validation in real quantum simulators and devices.
Therefore, progress in \gls{QSE} is essential, with the aim of establishing methodologies, languages and tools that allow the systematic and reliable engineering of quantum software \cite{ali2022software}.

%% file: sections/RelatedWork.tex
\section{Related Work} \label{sec:relatedwork}

Several secondary studies have explored the field of Quantum Software Engineering (QSE) from different perspectives. From \citet{DESTEFANO2024} have a systematic mapping on recurring topics and technologies used, while \citet{KHAN2023} perform a systematic review aimed at software architecture for quantum systems. \citet{Serrano2022} and \citet{Dwivedi2024} offer panoramic views on components, tools and the life cycle of quantum software development, and \citet{Hacaloglu2024} discuss the functional requirements and their measurement in the quantum context.
Other studies, such as \citet{AparicioMorales2024}, focus on the characterization of the Latin American scenario, with emphasis on collaborations, regional challenges and priority research lines. \cite{Piattini2024} and \citet{Dwivedi2024} also discuss trends and gaps in \gls{QSE} practice, based on broad surveys about the state of the art.

Despite the relevant contributions of these works, none of them focus on the proposition of a national research agenda, articulating specific strategic axes for the coordinated advance of \gls{QSE} in a country with low representativeness, as is the case of Brazil. This is the main differential of this study, which not only maps the state of the art of \gls{QSE}, but proposes a structured plan divided into four axes - education and training, interdisciplinary research groups, national scientific events and funding strategies (see Section \ref{sec:agenda}) - to foster the construction of an engaged and productive scientific community in Brazilian soil.

%% file: sections/ResearchMethod.tex
\section{Research Method} \label{sec:researchmethod}

This \gls{SMS} was organized into three main phases planning, conducting, and reporting, in accordance with the guidelines proposed by \cite{Kitchenham2007guidelines}. The following subsections describe the procedures adopted in each phase.

\subsection{Planning}
To guide the study, the following main research question was established: \textit{\rqzero}?
In this context, a set of secondary research questions was also defined to support data extraction and analysis:

\begin{itemize}
\item \textbf{\rqoneonecode}: \rqoneone 
\item \textbf{\rqonetwocode}: \rqonetwo 
\item \textbf{\rqonethreecode}: \rqonethree 
\item \textbf{\rqonefourcode}: \rqonefour 
\item \textbf{\rqonefivecode}: \rqonefive 
\item \textbf{\rqtwoonecode}: \rqtwoone 
\end{itemize}

These questions aim to uncover patterns in the temporal evolution of the field (RQ1), identify key contributors (RQ2 and RQ3), map the main publication vehicles (RQ4), and classify the methodological approaches employed (RQ5). Additionally, by aligning the analyzed studies with the SWEBOK knowledge areas (RQ6), the study provides insights into which \gls{QSE} domains are currently being explored within the quantum context.

In this study, ACM Digital Library, IEEE Digital Library, Scopus, and Web of Science databases were selected due to extensive coverage of peer-reviewed literature.
To ensure a comprehensive search, the search string was formulated based on the PICO framework, focusing on the Population and Intervention elements for being a \gls{SMS}, as recommended by \citet{Kitchenham2007guidelines}. Considering \gls{QSE} an adaptation of traditional \gls{SE} based on \gls{QC} fundamentals, we defined \gls{QC} synonymous representing the population and \gls{SWEBOK} areas as intervention related to \gls{SE}.
This approach was employed to retrieve studies addressing \gls{QSE} context. The final search string is presented in Table~\ref{tab:searchstring}.

\begin{table*}[!h]
    \footnotesize
    \centering
    \begin{tabularx}{\linewidth}{
        >{\RaggedRight\arraybackslash}p{2cm}
        >{\RaggedRight\arraybackslash}X    
        >{\RaggedRight\arraybackslash}p{1cm}    
    }
        \hline
        \textbf{Sets} & \textbf{Keywords} & \textbf{Logic} \\ \hline
        Population  & quantum OR ``quantum computing'' OR ``quantum software'' OR ``quantum program'' & AND \\
        Intervention    & ``software engineering'' OR ``software requirement'' OR ``software architecture'' OR ``software design'' OR ``software construction'' OR ``software test'' OR ``software engineering operation'' OR ``software maintenance'' OR ``software configuration management'' OR ``software engineering management'' OR ``software engineering process'' OR ``software engineering model'' OR ``software engineering methods'' OR ``software quality'' OR ``software security'' OR ``software engineering professional practice'' OR ``software engineering economic'' OR ``computing foundation'' OR ``mathematical foundation'' OR ``engineering foundation'' & ~ \\ \hline
    \end{tabularx}
    \caption{Search string applied}
    \label{tab:searchstring}
\end{table*}

The search was performed by applying the search string to the metadata (i.e., title, abstract, and keywords) of each article across the source. 
Inclusion and exclusion criteria were defined to filter publications relevant to the study. Studies that met all three inclusion criteria were selected and those that met at least one exclusion criterion were rejected. 
Inclusion criteria are
    (i) study must be in the \gls{QSE} context,
    (ii) publication must be from the latest study, and
    (iii) it published in conference, journal or book chapters.
Exclusion criteria are
    (i) publications as secondary studies,
    (ii) publications that can not be accessed completely, and
    (iii) publications that do not address aspects related to software production for the quantum computing paradigm (e.g., quantum computing algorithms, quantum mechanics, quantum information).


\subsection{Conducting}
The application of the review protocol yielded the following results (see Table \ref{tab:selectionprocess}).
Between February and April 2025, \totalretrieved~ results were found in the databases.
From these articles, the screening process began. 
Following inclusion and exclusion criteria while reading the title, abstract and full text, \totalexcluded~ studies were excluded, resulting in \totalexcluded~ articles for data extraction.
Parsifal\footnote{https://parsif.al} was used to support the screening process, Microsoft Excel to tabulate and export the data to be used in the analysis and generation of graphs in Google Colab\footnote{https://colab.google/}.

Due to the number of selected articles, we have made supplementary material available\footnote{\href{\repository}{\repository}} (e.g., full list of selected studies, scripts used to generate graphs, among others) that can be used to aggregate in the analysis of the main results described in Section \ref{sec:results} and Section \ref{sec:discussion}.


\begin{table}[ht]
    \footnotesize
    \begin{center}
        \begin{tabular}{ l c c c c }
        \hline
            \textbf{Source} & \textbf{\#retrieved} & \textbf{\#excluded}  & \textbf{\#final} \\
                            & \textbf{studies} & \textbf{studies} & \textbf{studies}\\
            \hline
            ACM Digital Library     & \acmretrieved & \acmexcluded & \acmaccepted \\
            IEEE Digital Library    & \ieeeretrieved & \ieeeexcluded & \ieeeaccepted \\
            Scopus                  & \wosretrieved & \wosexcluded & \wosaccepted \\
            Web of Science          & \scpretrieved & \scpexcluded & \scpaccepted \\
            \hline
            Total                   & \totalretrieved & \totalexcluded & \totalfinalstudies \\
            \hline
        \end{tabular}
        \caption{Studies selected and included}
        \label{tab:selectionprocess}
    \end{center}
\end{table}


%% file: sections/Results.tex
\section{Results} \label{sec:results}

\subsection{\rqoneone\ (\rqoneonecode)} \label{sec:results-year}
\begin{figure}[!h]
    \centering
    \includegraphics[width=1\linewidth]{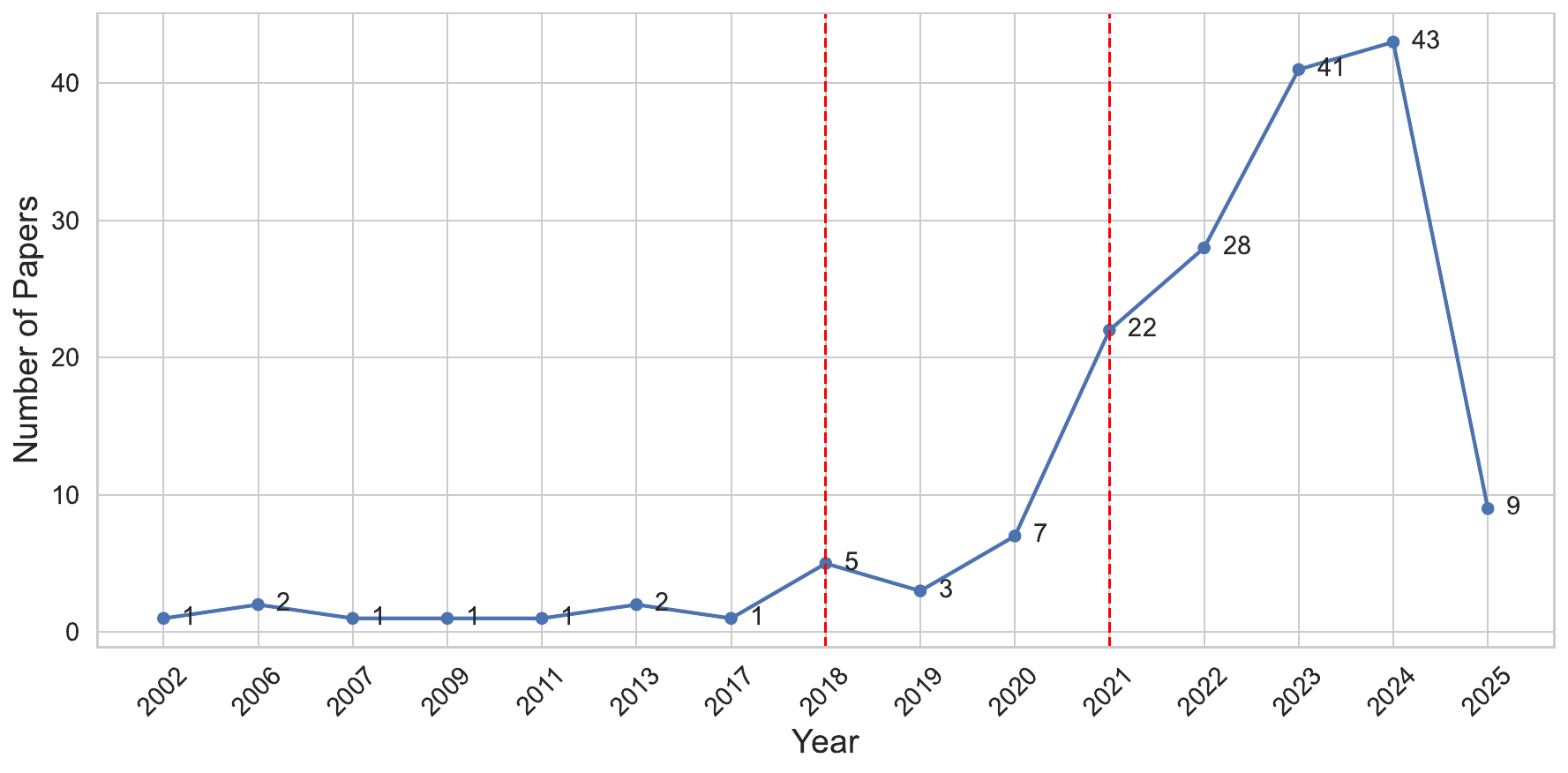}
    \caption{Number of publications per year}
    \label{fig:chart_year}
\end{figure}

To answer the first research question of this study, we analyzed the evolution in the number of studies related to \gls{QSE} published over the years. 
According to Figure~\ref{fig:chart_year}, we identified three key periods of publications. In the first period, the amount of publications are not relevant between 2002 and 2017, indicating that \gls{QSE} research was incipient. In the second period, the publications started a moderate growing between 2018 and 2020. Finally, in the third period, we can see an expressive increase in publications from one year to the next, i.e., each year from 2021, the number of works in \gls{QSE} has increased, arriving in 2024 with the historical maximum of 43 publications.

As presented in Section \ref{sec:researchmethod}, the search for studies was performed between February and April 2025, which justifies the number of papers published. However, the number of publications in this period exceeds 2020, indicating an increasing number of works by the end of 2025.

\subsection{\rqonetwo\ (\rqonetwocode)} \label{sec:results-institutions-authors}
\begin{figure}[!h]
    \centering
    \includegraphics[width=1\linewidth]{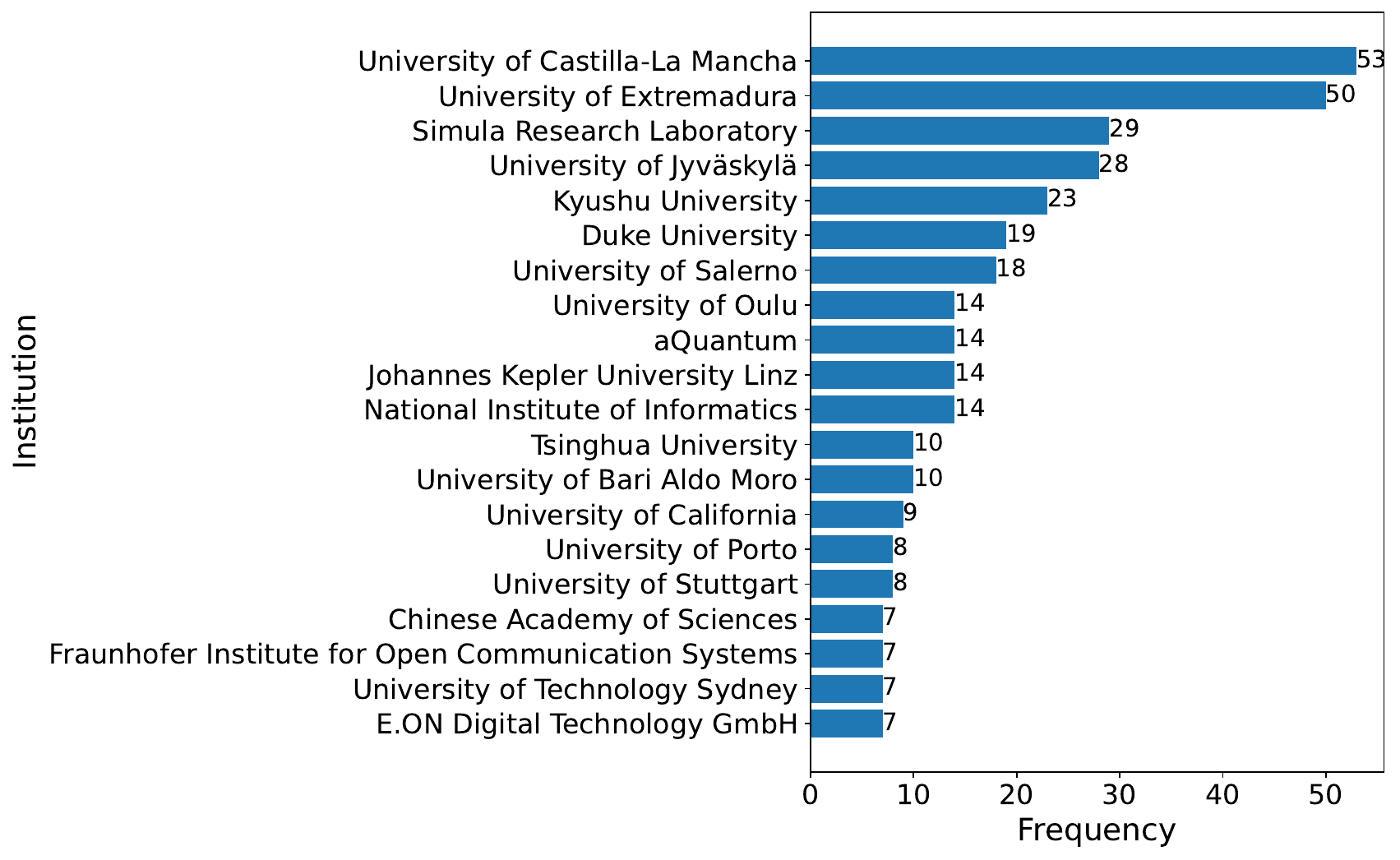}
    \caption{Top 20 most frequent institutions}
    \label{fig:chart_institutions}
\end{figure}

\begin{figure}[!h]
    \centering
    \includegraphics[width=1\linewidth]{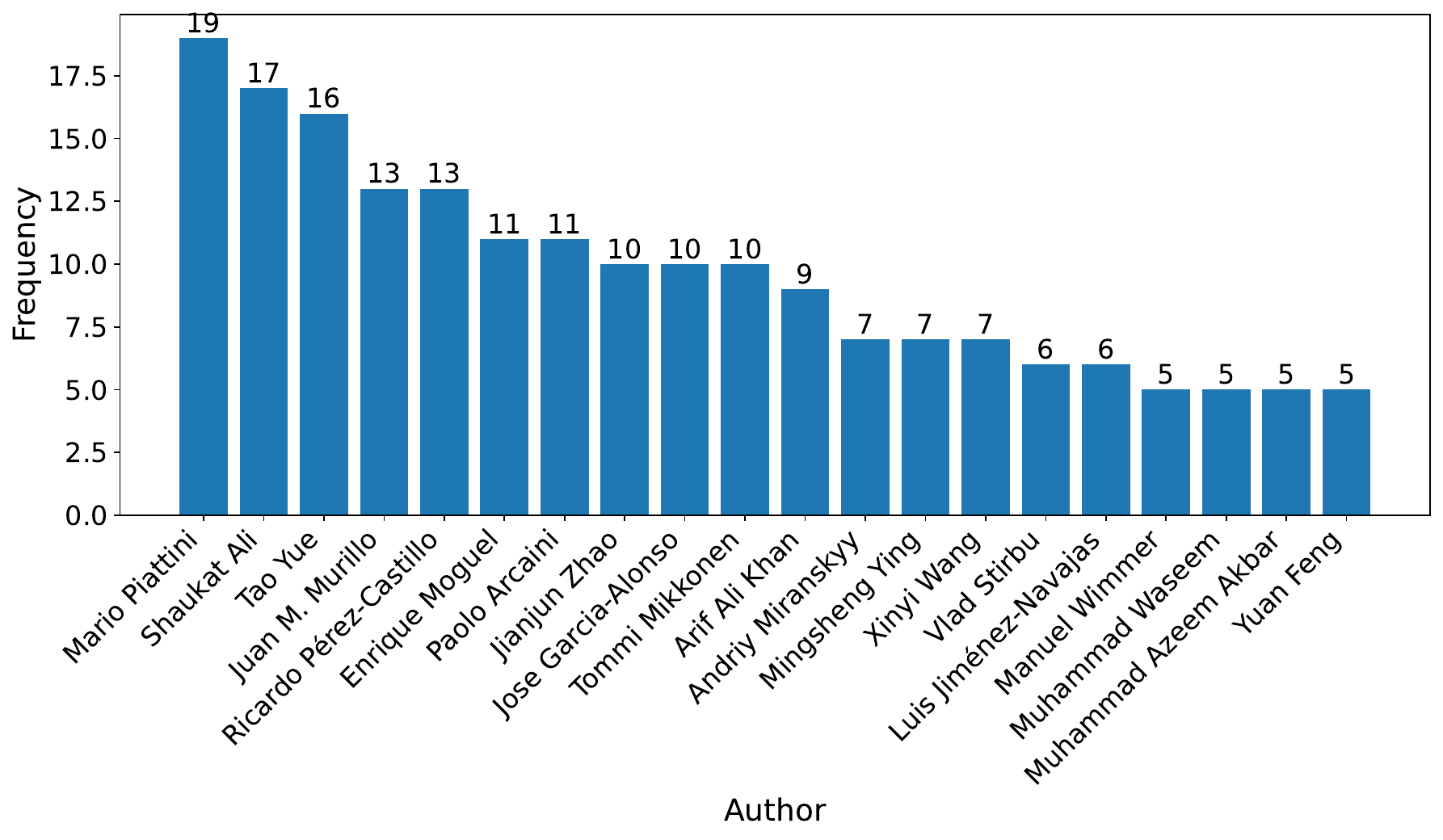}
    \caption{Top 20 most frequent authors}
    \label{fig:chart_authors}
\end{figure}

In this research question, we are interested in identifying the main institutions and authors that have contributed to the advancement of \gls{QSE}.
According to Figure~\ref{fig:chart_institutions}, the institutions that stand out are \textit{University of Castilla-La Mancha} (53) and \textit{University of Extremadura} (50), both Spanish institutions. In the case of \textit{University of Castilla-La Mancha}, this result can be associated with the leading role of researcher \textit{Mario Piattini} (19), as can be seen in Figure~\ref{fig:chart_authors}.

It is noted that the publication indicators of the institutions are not directly influenced by their associated researchers. This is the case of researcher \textit{Shaukat Ali} (17), who stands out as the second most influential researcher in \gls{QSE}. However, the institution to which he is associated (\textit{Simula Reseach Laboratory}) is third in the ranking of institutions, with 29 publications.


\subsection{\rqonethree\ (\rqonethreecode)} \label{sec:results-country}

\begin{figure*}
    \centering
    \includegraphics[width=1\linewidth]{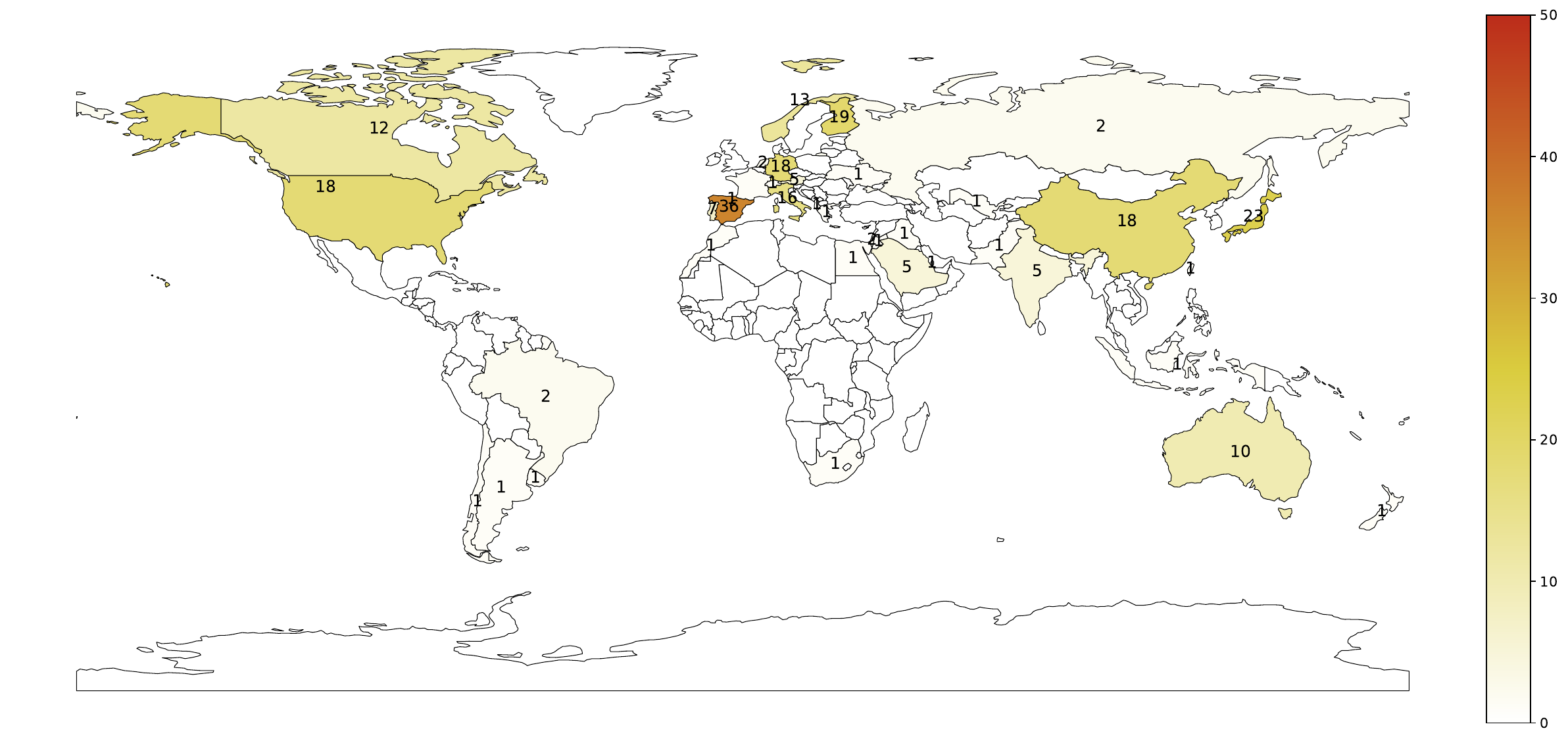}
    \caption{Global distribution of publications by country}
    \label{fig:chart_worldmap}
\end{figure*}

Figure~\ref{fig:chart_worldmap} shows the global distribution of scientific production in \gls{QSE}, highlighting the countries that contributed most publications in the area. The graph shows that there is a concentration of publications in certain regions, while others remain insignificant or absent. The intensity of the colors on the map allows us to identify the countries with the largest number of articles, offering an overview of the international insertion of the area.

Table~\ref{tab:continent_country_articles} complements Figure~\ref{fig:chart_worldmap} by showing the 20 countries with the highest number of publications, grouped by continent. The leader is \textit{Spain} (36), followed by \textit{Japan} (23), \textit{Finland} (19), as well as \textit{China}, \textit{Germany} and \textit{United States}, both with 18 publications. 
This indicates that most of the production is concentrated in developed countries.

When grouping countries by continent, it can be seen that Europe is the main hub of contributions in \gls{QSE}, bringing together more than half of the countries listed. Asia also demonstrates relevance. In the Americas, although the \textit{United States} and \textit{Canada} are among the most productive, \textit{Brazil} appears with low participation (2). Oceania is represented only by Australia, with 10 publications.

\begin{table}[ht]
\footnotesize
\centering
\caption{Distribution of top 20 countries by continent}
\begin{tabular}{llc}
\hline
\textbf{Continent} & \textbf{Country} & \textbf{\#Article} \\
\hline
Europe        & Spain           & 36 \\
        & Finland         & 19 \\
        & Germany         & 18 \\
        & Italy           & 16 \\
        & Norway          & 13 \\
        & Portugal        & 7 \\
        & Austria         & 5 \\
        & United Kingdom  & 4 \\
        & Netherlands     & 2 \\
\hline
Asia          & Japan           & 23 \\
          & China           & 18 \\
          & India           & 5 \\
          & Saudi Arabia    & 5 \\
          & Russia          & 3 \\
          & South Korea     & 3 \\
          & Israel          & 2 \\
\hline
America       & United States   & 18 \\
       & Canada       & 12 \\
       & Brazil          & 2 \\
\hline
Oceania       & Australia       & 10 \\
\hline
\end{tabular}
\label{tab:continent_country_articles}
\end{table}

\subsection{\rqonefour\ (\rqonefourcode)}
\begin{figure*}[!t]
    \centering
    \includegraphics[width=\linewidth]{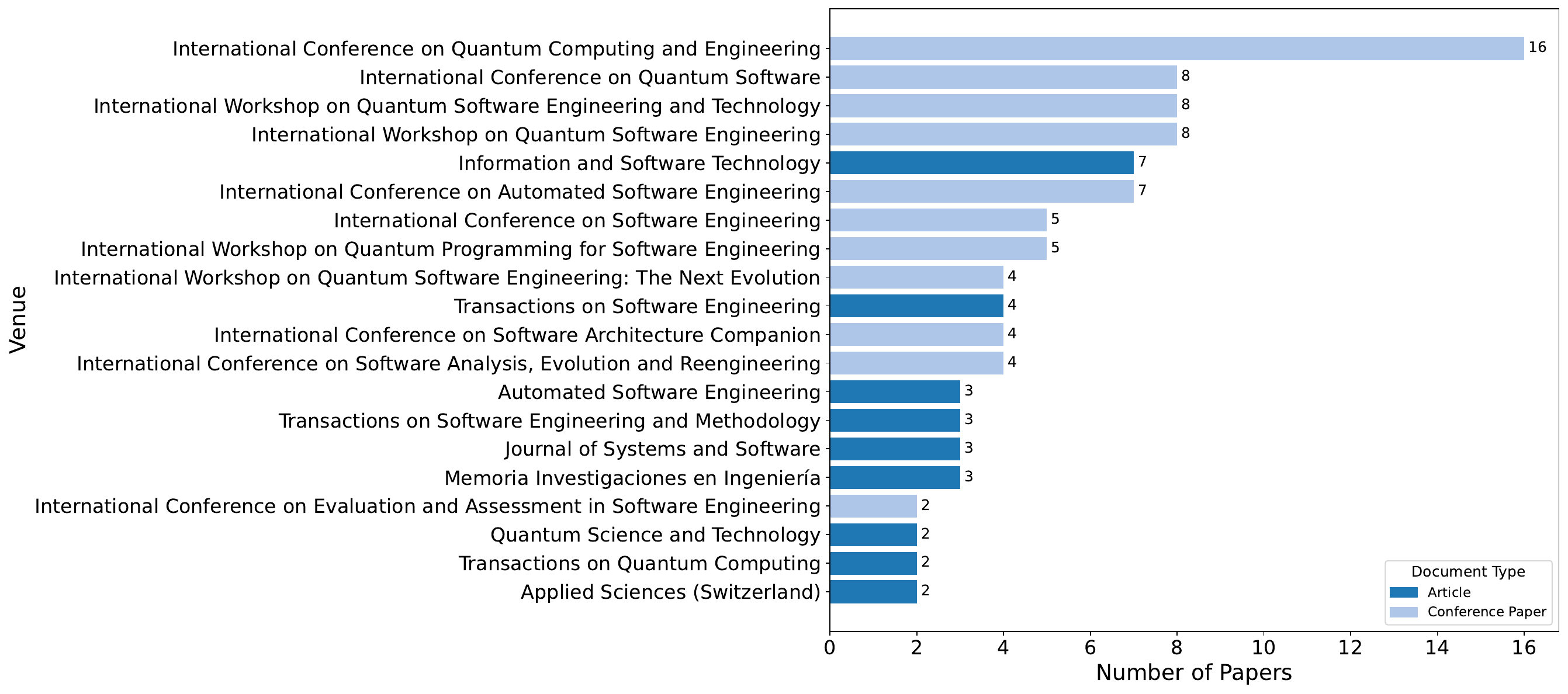}
    \caption{Top 20 venues by number of papers, grouped by document type}
    \label{fig:chart_venues}
\end{figure*}

The answer to this research question is very relevant to the scientific community, as it provides an overview of the main publication vehicles, in addition to helping in understanding how the community is structuring and sharing the results and contributions in \gls{QSE}.
Figure~\ref{fig:chart_venues} presents the ranking of the top 20 publication outlets and the relationship between the type of papers, i.e., \textit{article} for journals and \textit{conference paper} for conferences (data obtained from the metadata of the search databases).

From our analysis, we identified that most of the studies were published in conferences, with emphasis on the
\textit{International Conference on Quantum Computing and Engineering} (16), followed by the
\textit{International Conference on Quantum Software}, 
\textit{International Workshop on Quantum Software Engineering and Technology}, 
\textit{International Workshop on Quantum Software Engineering},
all conferences with 8 published papers.
\textit{Information and Software Technology} is the scientific journal that stands out from the others with 7 publications.
In general, venues such as \textit{Information and Software Technology} and \textit{Transactions on Software Engineering} are well-established journals, indicating that part of the research in \gls{QSE} is being absorbed by traditional \gls{SE} literature.

\begin{figure}[h]
    \centering
    \includegraphics[width=.6\columnwidth]{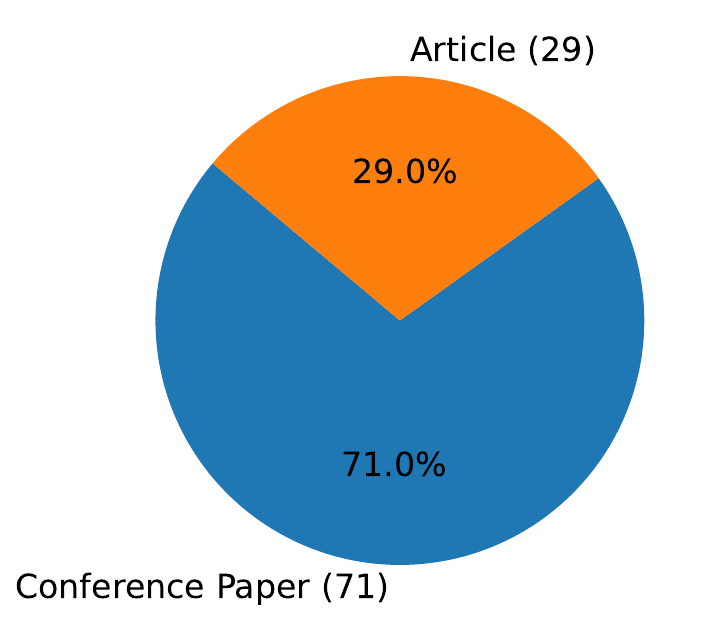}
    \caption{Distribution of document types among the top 20 venues}
    \label{fig:chart_documenttype}
\end{figure}

We concluded that the majority of the publications in the top venues are conference papers (75), compared to a smaller but relevant number of journal articles (29), as shown in Figure~\ref{fig:chart_documenttype}. The dominance of conference papers is typical of fast-developing areas, where rapid feedback and community interaction are essential.

\subsection{\rqonefive\ (\rqonefivecode)} \label{sec:results-study-types}

\begin{figure}
    \centering
    \includegraphics[width=1\linewidth]{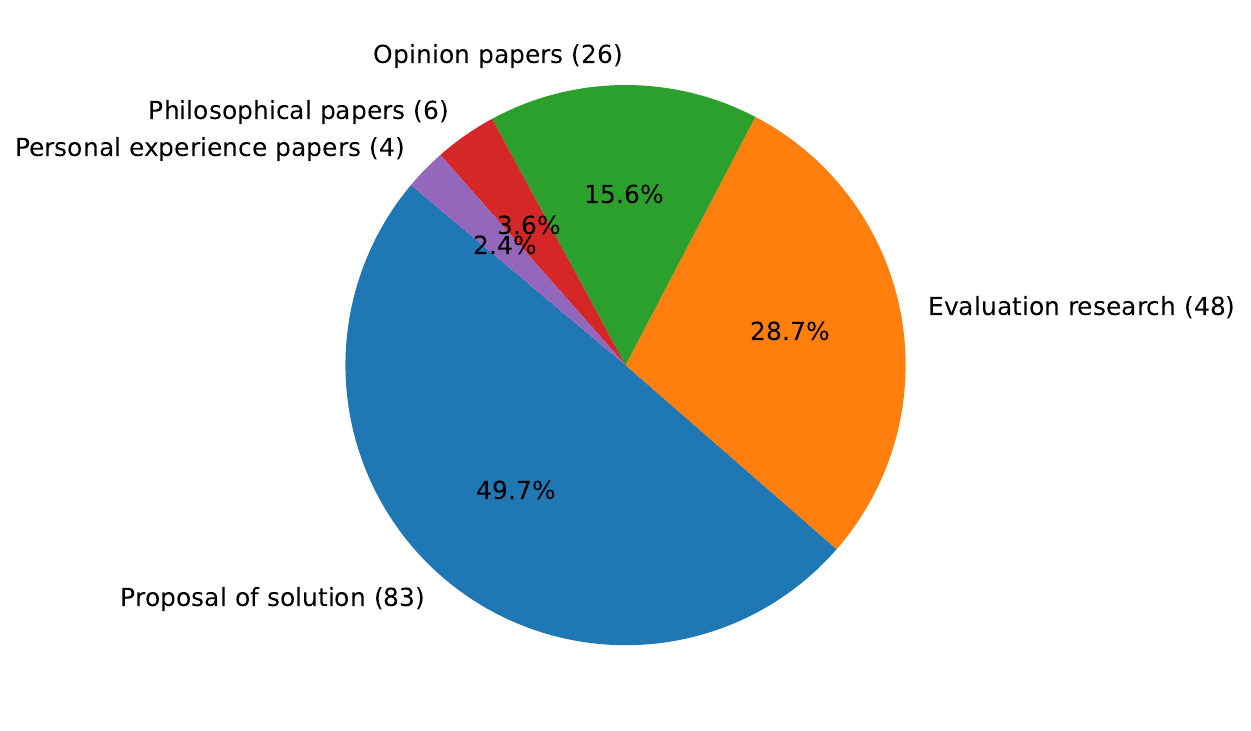}
    \caption{Distribution of research types among primary studies}
    \label{fig:chart_researchtype_pie}
\end{figure}

The aim of this research question is to 
investigate the predominant methodological approaches among primary studies.
These studies were classified according to the six well-established research types proposed by \citet{Wieringa2006}, namely: evaluation research, proposal of solution, validation research, philosophical papers, opinion papers, and personal experience papers.
According to Figure~\ref{fig:chart_researchtype_pie}, primary studies focus mainly on proposed solutions (83), followed by evaluation research (48). Opinion articles (26) were also identified, in addition to a small number of philosophical articles (6) and reports of personal experience (4).
These results demonstrate that the community is strongly focused on proposing solutions to problems that are still open in \gls{QSE}. Furthermore, the expressiveness of evaluation research type works indicates that the community is concerned with validating the proposed solutions through scientific methods consolidated in the literature. On the other hand, the lower occurrence of articles in the other categories shows that more theoretical, critical or experience-based reflections still occupy a secondary position in the current panorama, but that they are also essential for the advancement of the area.


\subsection{\rqtwoone\ (\rqtwoonecode)} \label{sec:results-swebok}

In order to understand the scope of contributions in \gls{QSE}, each study was classified into one of the 19 areas of \gls{SWEBOK} \cite{swebok2024}. The categorization was carried out exclusively, that is, associating each article with a single main area, according to the predominant scope of the study. This decision was made to facilitate the analysis and main contributions of the studies.
In cases where the contribution was broader, without a direct connection to a specific area of \gls{SWEBOK}, the additional category called \textit{The Quantum Software Engineering in General} was adopted. Most of the articles in this category are reports that provide an overview of quantum computing, state of the art and challenges for \gls{QSE}.

\begin{figure}[h]
    \centering
    \includegraphics[width=1\linewidth]{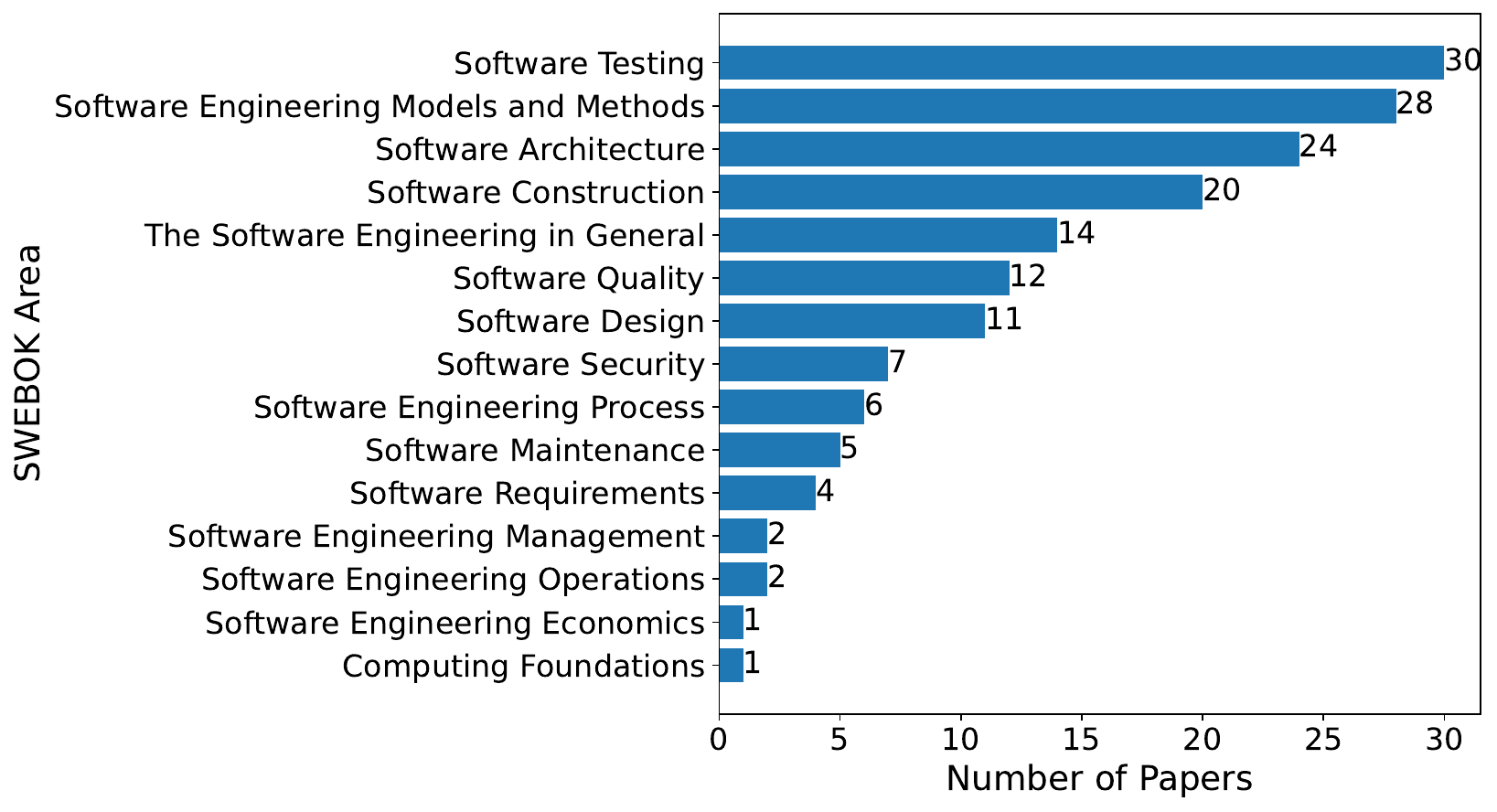}
    \caption{Distribution of \gls{SWEBOK} areas addressed in the selected studies}
    \label{fig:chart_swebok}
\end{figure}

As a result of our analysis, the most frequently addressed areas were \textit{Software Testing} (30), \textit{Software Engineering Models and Methods} (28) and \textit{Software Architecture} (24), as shown in Figure~\ref{fig:chart_swebok}.
This indicates that the community has been working hard to ensure the quality of quantum software development, by focusing studies on the development of tests; techniques, methods and tools to support development; and the structure by which quantum software and services can be structured.
We understand that the recurrence of these categories suggests that the community has focused on technical and structural aspects of \gls{SE}, adapting classical concepts to the peculiarities of quantum computing.

%% file: sections/Discussion.tex
\section{Discussion} \label{sec:discussion}

\subsection{Temporal Trends and Venues}

\begin{figure}
    \centering
    \includegraphics[width=1\linewidth]{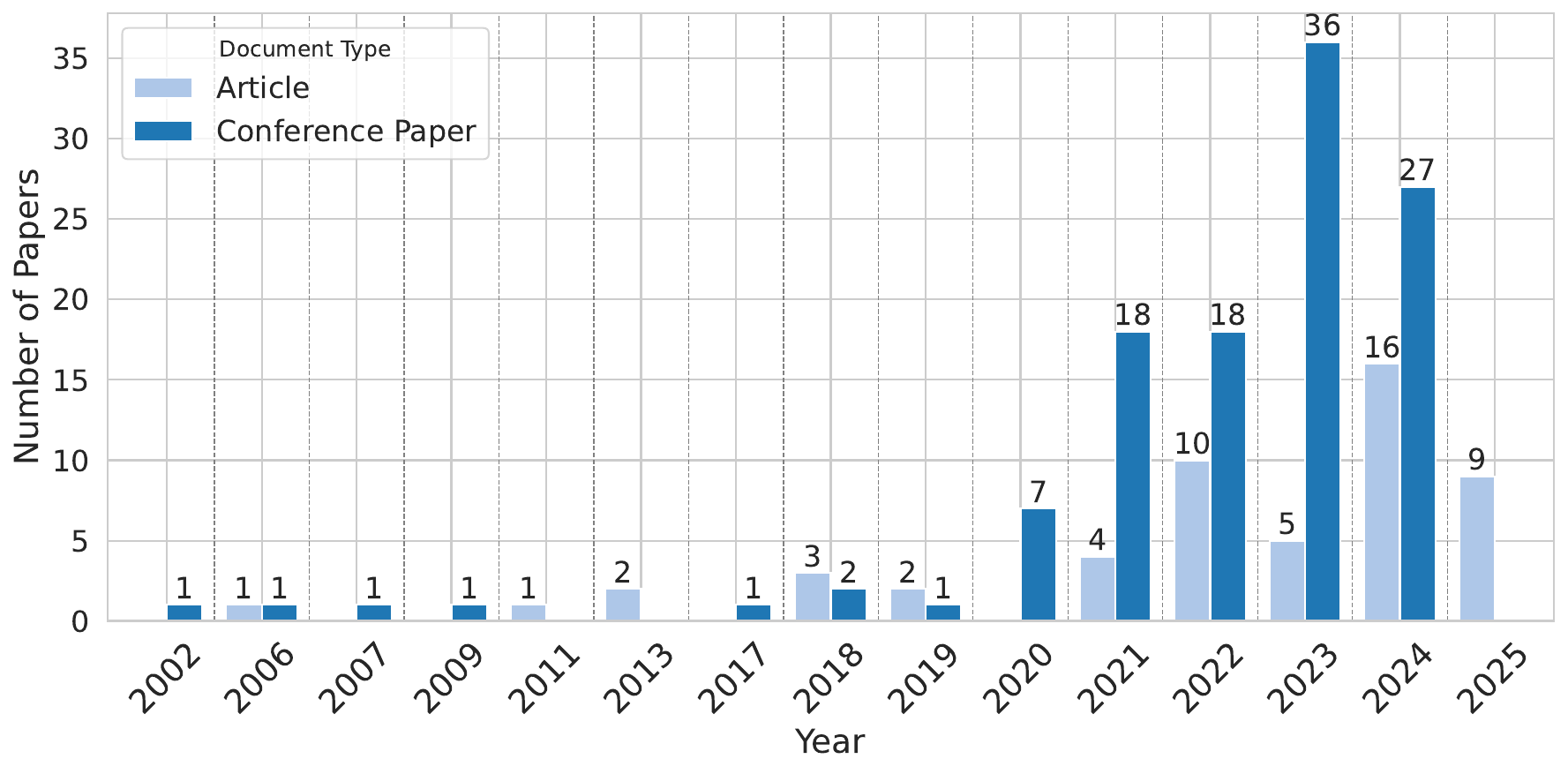}
    \caption{Evolution of document types over time}
    \label{fig:chart_year_doctype}
\end{figure}

As shown in Figure~\ref{fig:chart_year}, there has been a significant growth in publications since 2021. In order to better understand this behavior, we categorized the types of documents published over the years, as shown in Figure~\ref{fig:chart_year_doctype}.

Until 2019, there was little variation between conference and journal publications. Since 2020, conference publications have consistently outnumbered journal publications in every year. This predominance can be explained by the shorter peer-review time typical of conferences, which facilitates the rapid dissemination of results in emerging areas such as \gls{QSE}. However, although more numerous, conference papers tend to present more technical and specific contributions, while journal papers generally have greater methodological depth and longer-lasting scientific impact.

A relevant development of this temporal analysis is the verification of the history of publication vehicles. It is possible that the significant increase in the number of publications in conferences since 2020 is directly related to the creation of specialized forums on \gls{QSE}.
Events such as the
\textit{International Conference on Quantum Computing and Engineering} and \textit{International Workshop on Quantum Software Engineering and Technology}, both started in 2020,
the \textit{International Workshop on Quantum Software Engineering}, launched in 2021,
and the \textit{International Conference on Quantum Software} (QSW), created in 2022,
have started to offer spaces dedicated to technical, conceptual and experimental discussions in the area. The emergence of these forums not only expanded publication opportunities, but also favored the consolidation of a scientific community around the subject.

\subsection{SWEBOK Focus by Research Strategy}

Understanding how research types have been applied in different areas of \gls{QSE} is essential to assessing their stage of maturity. By relating them to the areas of \gls{SWEBOK} (Figure~\ref{fig:chart_swebok}), it is possible to identify which aspects have received greater attention, how this has manifested itself methodologically, and where there are still important gaps. 

\begin{figure*}
    \centering
    \includegraphics[width=.8\linewidth]{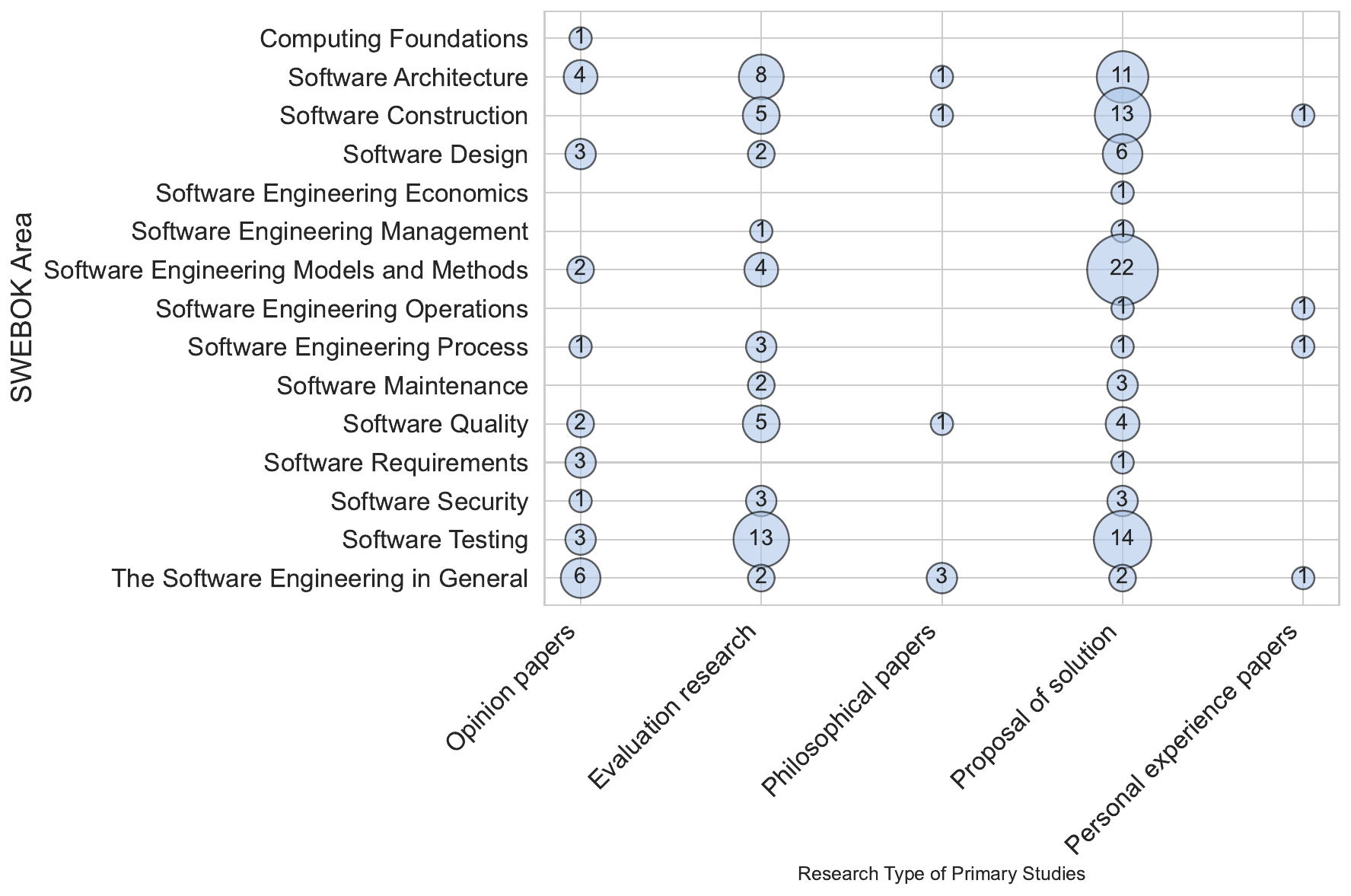}
    \caption{Relationship between research strategies and SWEBOK areas among primary studies}
    \label{fig:chart_swebok_researchtype}
\end{figure*}

Figure~\ref{fig:chart_swebok_researchtype} summarizes the relationship between the types of research conducted in the primary studies and the \gls{SWEBOK} domains addressed by these works.
The areas of
\textit{Software Engineering Models and Methods},
\textit{Software Testing},
\textit{Software Construction} and
\textit{Software Architecture}
concentrate most of the studies classified as \textit{Proposal of solution}.
This suggests that the papers introduce a novel technique or a substantial enhancement of an existing one, and justify its relevance, even though they do not provide comprehensive validation. The proposed approach is typically supported by a proof of concept, which may take the form of a simplified example, a logical argument, or another illustrative means.

Furthermore, \textit{Evaluation research} is also significant in \textit{Software Testing},
\textit{Software Architecture},
\textit{Software Construction}, and 
\textit{Software Quality}.
In this type of research, there is a clearer concern with the empirical evaluation of techniques, tools or models applied to real or simulated contexts. The significant presence of this type of study in these categories may indicate a greater degree of methodological maturity, with an emphasis on verifying the practical effectiveness of the proposed solutions, which contributes to strengthening the body of knowledge of \gls{QSE} based on concrete evidence.
The prominence of these areas can be understood in light of the fundamental challenges imposed by the quantum paradigm on the software development process, such as probabilities, non-determinism, entanglement and superposition of states.

In the context of \textit{Software Testing}, the impossibility of directly observing quantum states without collapsing them requires testing strategies that use statistical sampling, repeated testing, and validation by behavioral equivalence.
The area of \textit{Software Quality}, on the other hand, demands the reformulation of traditional metrics, incorporating attributes such as fidelity, coherence, noise, and robustness to decoherence.
With regard to \textit{Software Architecture}, it is essential to structure hybrid applications — which integrate classical and quantum components — respecting physical, temporal, and orchestration constraints between different computing environments.
In addition, since access to real quantum computers occurs exclusively through cloud services or APIs provided by specific platforms, it becomes even more necessary to develop robust software architectures capable of efficiently managing communication between the classical layers of the application and remote quantum backends.

Finally, in \textit{Models and Methods}, there is a growing emphasis on creating formal abstractions, \glspl{DSL} and visual notations that allow for the clear representation and facilitation of modeling of quantum operations and manipulation of quantum circuits.
These challenges highlight the need to advance not only in technical solutions, but also in conceptual frameworks that allow \gls{SE} to engage with the foundations of quantum computing.

\subsection{Geographic and Institutional}

The analysis of the geographic and institutional distribution of studies in \gls{QSE} can reveal important patterns regarding the consolidation and dissemination of this emerging area. By identifying the most productive institutions and their respective countries, it becomes possible to understand where the main research centers are located and which regions concentrate the greatest effort. This analysis is strategic for identifying academic leaders, potential collaboration networks and also reflecting on the scientific production related to \gls{QSE}.

\begin{figure*}[!h]
    \centering
    \includegraphics[width=.8\linewidth]{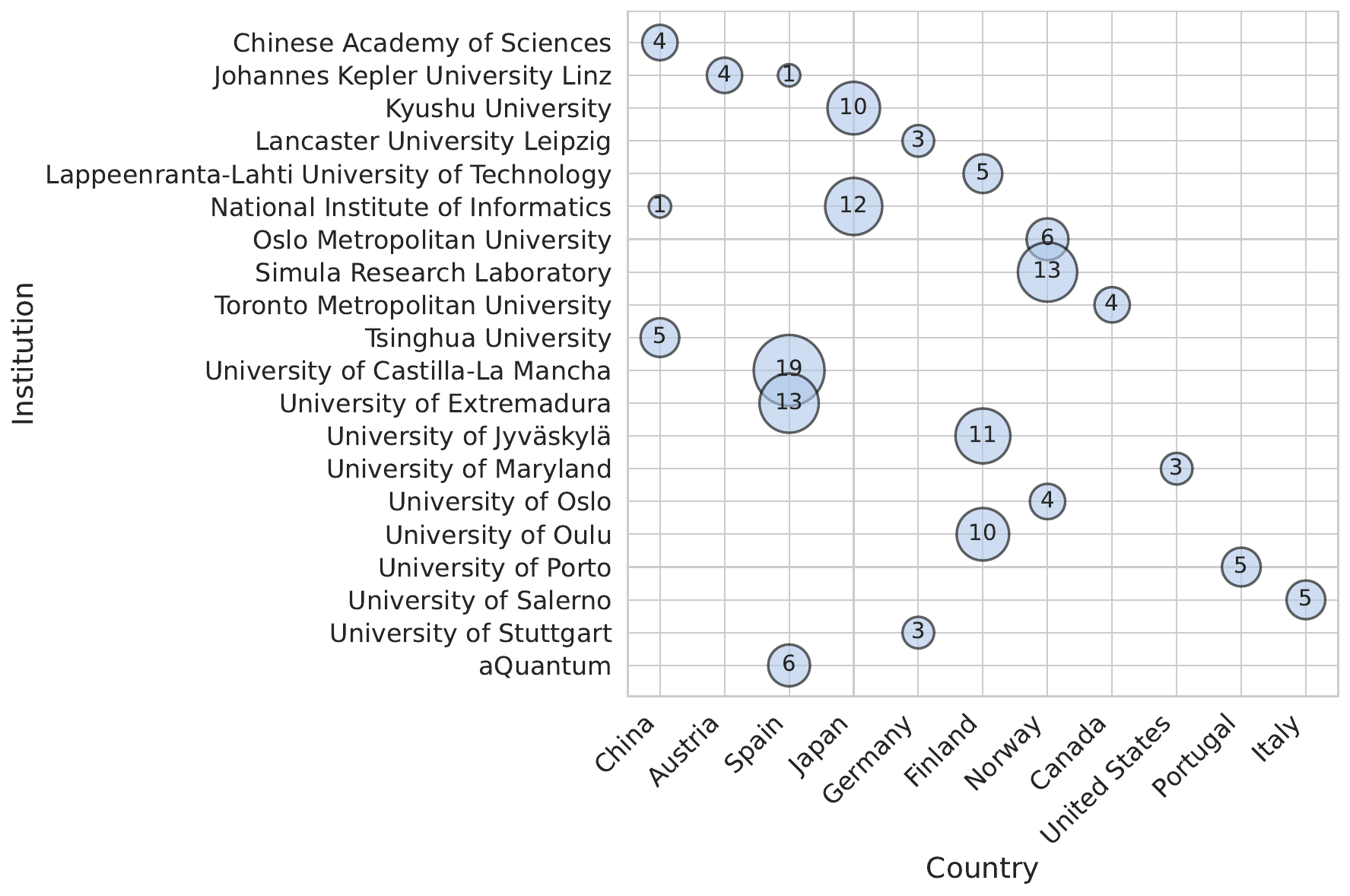}
    \caption{Top 20 Institutions by country and number of publications}
    \label{fig:chart_country_institution}
\end{figure*}

According to Figure~\ref{fig:chart_country_institution}, most institutions with significant production are located in Europe (Spain, Finland and Norway).
This may indicate that European research centers are mobilizing coordinated efforts to position themselves as leaders in \gls{QSE}. There is strong funding from European programs such as \textit{Horizon Europe}\footnote{\href{https://research-and-innovation.ec.europa.eu}{https://research-and-innovation.ec.europa.eu}}, in addition to international collaborations within the European Union that facilitate scientific exchange and shared funding.

Spain stands out in particular with three highly productive institutions, namely
\textit{University of Castilla-La Mancha} (19) and
\textit{University of Extremadura} (13), as well as 
\textit{aQuantum} (6) as a company specialized in research, development, consulting and services in the fields of \gls{QSE} and development.
Spain is not traditionally a global leader in quantum computing, but the data suggests the existence of groups strongly dedicated to \gls{QSE}, such as \textit{Mario Piattini},
\textit{Ricardo Pérez-Castillo} and
\textit{Enrique Moguel}, as shown in (Figure~\ref{fig:chart_authors}). 

We can also highlight Norway, which, despite its small territorial size and population, is one of the countries with the highest number of publications in \gls{QSE}. This leading role is largely due to the work of the \textit{Simula Research Laboratory}, which concentrates the largest number of articles in the country, and to researcher \textit{Shaukat Ali}, one of the most productive authors in the area. The combined presence of an institution with a strong focus on \gls{SE} and an influential researcher highlights the importance of highly specialized centers of excellence in advancing research in emerging domains.

Despite the prominence of Japan and China, Asian participation is limited. Brazil, for example, does not appear in the list of major institutions.
This scenario possibly indicates the lack of structured research programs in \gls{QSE}, the absence of consolidated groups in this sub-area and the recent entry of the topic into local scientific agendas.
The lack of funding and the difficulty of international coordination may also be limiting factors.

\subsection{Threats to Validity}


This study is subject to some threats to validity. Definition of search string and exclusive categorization by \gls{SWEBOK} area may have limited the scope of the captured studies. Although four consolidated bases have been used, other relevant sources Springer Link, Engineering Village or even arXiv have not been included, which may impact the representativeness of the results. Finally, the emerging nature of \gls{QSE} implies that the data analyzed, collected until the first trimester of 2025, can rapidly become outdated in the face of the acceleration of publications in the area.

%% file: sections/Agenda.tex


\section{Brazilian Research Agenda} \label{sec:agenda}

Despite the global growth of \gls{QSE}, led by countries such as Spain, Japan, Finland, China, Germany and the United States, Brazil’s involvement in the field is still incipient, with only 2 identified publications. This scenario highlights the urgent need for structured actions that boost the national presence in the area. Therefore, this work proposes a Brazilian research agenda in \gls{QSE}, with the objective of articulating training, collaboration and financing to stimulate scientific production and consolidate Brazil as a relevant actor in this strategic field. Agenda is organized on four main axes: \textit{education and training}, \textit{fostering interdisciplinary research groups}, \textit{establishing national events} and \textit{funding strategies}.



\subsection{Education and training}

In order for the \gls{QSE} field to expand consistently in Brazil, it is necessary to invest primarily in education. However, one of the greatest challenges — and at the same time a primary condition — is to teach, in an accessible and effective manner, the fundamental concepts of quantum mechanics and quantum computing \cite{TeachingQuantumComputing}. These are contents that require a high level of abstraction and break with many of the intuitions formed from classical logic, demanding new pedagogical and interdisciplinary approaches.

In view of this, it is urgent to gradually include content related to quantum computing in undergraduate and graduate courses in Computing, Software Engineering, and related areas. Specific disciplines, integrative projects, and extension initiatives can introduce concepts such as qubits, entanglement, gates, and quantum algorithms, in a way it is aligned with the skills already developed in the traditional curriculum. This curricularization will not only expand quantum literacy among future professionals, but will also create conditions for the emergence of new research groups and practical applications of \gls{SE} in the quantum context.

For example, several educational and corporate initiatives have been implemented in Brazil with the aim of expanding training in quantum computing and communication, demonstrating an important movement towards the consolidation of this area in the national context. Institutions such as \textit{SENAI CIMATEC University}\footnote{\href{https://www.universidadesenaicimatec.edu.br/}{https://www.universidadesenaicimatec.edu.br/}} have been standing out by offering free and pioneering specializations focused on both quantum computing and communication, covering everything from algorithms to cryptographic protocols. The State University of Campinas (UNICAMP)\footnote{\href{https://unicamp.br/en/}{https://unicamp.br/en/}}, through the Gleb Wataghin Institute of Physics (IFGW), and the Pontifical Catholic University of Rio de Janeiro (PUC-Rio)\footnote{\href{https://www.puc-rio.br/english/}{https://www.puc-rio.br/english/}}, with its course on fundamentals and applications of quantum computing, also actively contribute to the dissemination of knowledge at the extension level.

In the private sector, companies such as QuaTI\footnote{\href{https://www.quati.tech/}{https://www.quati.tech/}} have dedicated themselves to professional training through courses that address algorithms, optimization and quantum machine learning.

Although dispersed, these initiatives indicate significant progress in building skills and raising awareness of the challenges and opportunities of the quantum era. Integrating them into a national strategy for training in \gls{QSE}, with a focus on curricular inclusion, teacher training and encouraging scientific initiation, is a challenging but fundamental step to position Brazil more effectively in this emerging field.

\subsection{Fostering interdisciplinary research groups}

In addition to efforts focused on education and training, the advancement of \gls{QSE} in Brazil also depends on the creation of a collaborative interinstitutional and interdisciplinary research network. The nature of quantum computing requires constant dialogue between different areas of knowledge, such as quantum mechanics, algorithm optimization, computer science, experimental physics, applied mathematics and, finally, \gls{QSE} itself. In this ecosystem, \gls{QSE} emerges as a point of convergence between theory and practice, requiring approaches that integrate formalism, abstraction, simulation, tool development and empirical validation.


Some initiatives have been consolidating research groups dedicated to quantum technologies, such as the Quantum Optics and Information Group\footnote{\href{http://sites.if.ufrj.br/qoqi/}{http://sites.if.ufrj.br/qoqi/}} from the Federal University of Rio de Janeiro (UFRJ), the Quantum Optics and Quantum Information Group \footnote{\href{https://www.fisica.ufmg.br/laboratorios/grupo-de-otica-quantica-e-informacao-quantica/}{https://www.fisica.ufmg.br/laboratorios/grupo-de-otica-quantica-e-informacao-quantica/}} from the Federal University of Minas Gerais (UFMG),
the Quantum Computing Group from Federal University of Santa Catarina \footnote{\href{http://www.gcq.ufsc.br/doku.php}{http://www.gcq.ufsc.br/doku.php}}, 
the Quanta Group\footnote{\href{https://portal.if.usp.br/pesquisa/pt-br/node/383}{https://portal.if.usp.br/pesquisa/pt-br/node/383}} from the Institute of Physics of the University of São Paulo (IFUSP), the National Laboratory of Scientific Computation or the National Laboratory for Scientific Computing (LNCC) \footnote{\href{https://www.gov.br/lncc/pt-br}{https://www.gov.br/lncc/pt-br}}, the Rio Quantum Network\footnote{\href{https://www.uff.br/16-09-2024/rede-de-internet-quantica-integra-instituicoes-de-ensino-e-pesquisa-no-rio/}{https://www.uff.br/16-09-2024/rede-de-internet-quantica-integra-instituicoes-de-ensino-e-pesquisa-no-rio/}} from Fluminense Federal University (UFF), among others.

These efforts, for the most part, focus on physical, theoretical and infrastructure aspects of quantum computing and communication. However, for the community to advance in a structured and strategic manner, it is essential that interdisciplinary research groups focused specifically on \gls{QSE} be formed, capable of integrating expertise in physics, computing, mathematics and \gls{SE} to deal with the challenges of developing reliable, scalable and reproducible quantum systems.
Therefore, creating research centers in \gls{QSE} that articulate these complementary skills can be an essential strategy to accelerate national scientific production and achieve significant contributions to the area.

\subsection{Establishing national events}

According to the analysis carried out in Section \ref{sec:results}, the publication venues revealed that most of the research in \gls{QSE} was published in recent and emerging international conferences, particularly since 2020.
Although international forums have played a crucial role in the advancement of \gls{QSE}, Brazil still lacks dedicated spaces where researchers, students and professionals can collectively engage in discussions and collaborations focused on the area.

Therefore, the establishment of national events — such as workshops, symposiums or dedicated tracks within existing conferences — represents a strategic step to promote knowledge exchange, encourage interdisciplinary initiatives and stimulate contributions from across the country. These forums can also serve as a kind of link between research groups and future international collaborations, strengthening Brazil's presence in the global \gls{QSE} scenario.

Although they are not focused on \gls{QSE}, Brazil already has some places where the scientific community can contribute to Quantum Computing and related topics, such as the Quantum Networks Workshop (WQuNets) and the Workshop on Quantum Communication and Computing (WQuantum). These initiatives represent a possibility of gradual integration with \gls{QSE}.

Furthermore, by positioning \gls{QSE} as an emerging track within these forums, a fertile environment is created to attract new researchers and spark the interest of students, contributing to the formation of a critical mass in the country.

\subsection{Funding strategies}

The advancement of \gls{QSE} in Brazil requires much more than good intentions and well-structured proposals — it depends, above all, on consistent financing strategies.
It is not enough to invest in education, form research centers or create discussion forums if these initiatives are not supported by adequate and continuous financial resources.

Funding is the link that makes all the other actions proposed in this agenda possible: from funding scholarships, acquiring infrastructure (which is very expensive), participating in scientific events and publishing results, to maintaining collaborative networks and developing specific tools for the quantum context. Without this support, even the most promising ideas are lost due to the lack of concrete conditions for their implementation.

Fortunately, Brazil has demonstrated significant efforts in funding research in quantum technologies through strategic initiatives. Of note are FAPESP's QuTIa program and the investment by the Ministry of Science, Technology and Innovation (MCTI) in the Embrapii Competence Center in Quantum Technologies at SENAI CIMATEC. Institutional partnerships, such as the one between Banco Inter\footnote{\href{https://inter.co/}{https://inter.co/}} and UFMG, also reinforce the engagement of the private sector. Although still focused on quantum hardware and communication, these initiatives signal an environment conducive to expanding investment in complementary areas such as \gls{QSE}, which requires specific support to grow in an integrated and interdisciplinary manner.

Therefore, recognizing the central role of funding is a fundamental step to ensure that \gls{QSE} continues to be a solid and transformative research front in the country.

%% file: sections/Conclusion.tex
\section{Conclusion} \label{sec:conclusion}



Our findings revealed a significant evolution of \gls{QSE} in recent years, with a significant increase in publications since 2021, a strong geographic concentration in European and Asian countries, leadership of authors and institutions, a predominance of primary studies aimed at proposing solutions that are still poorly validated, and a focus on the areas of \textit{Software Testing}, \textit{Software Engineering Models and Methods} and \textit{Software Architecture} of SWEBOK. Furthermore, we identified that the presence of the Brazilian scientific community is still incipient, which reinforces the urgency of actions to expand scientific and technological activity in \gls{QSE}.


Findings of this review supported the construction of a research agenda focused on the reality of Brazilian research in \gls{QSE}, in order to guide investments, articulate the scientific community and promote the consolidation of the country as a relevant actor in the evolution of \gls{QSE}.
The proposed agenda includes four main axes: (i) strengthening education and training in quantum computing and \gls{QSE}; (ii) formation of interdisciplinary research centers; (iii) creation of national scientific events dedicated to the topic, stimulating scientific production and aligned with international challenges; and (iv) definition of continuous funding strategies.

Finally, it is expected that the research agenda in this work can be quickly incorporated by researchers, institutions, and funding agencies. The consolidation of this agenda will not only expand Brazil's scientific presence in the area, but will also foster interdisciplinary collaborations, the training of qualified human resources and the development of technologies aligned with the challenges of the quantum paradigm.